\documentclass[10pt, conference, compsocconf]{IEEEtran}
\usepackage[pdftex]{graphicx}
\usepackage{times}
\usepackage{epstopdf}
\usepackage{amsfonts}
\newcommand{\F}{\mathbb{F}}
\usepackage{graphicx}
\PrependGraphicsExtensions*{.png,.PNG}
\newtheorem{exmp}{Example}
\renewcommand{\thetable}{\arabic{table}}
\ifCLASSINFOpdf
\else
\fi

\begin{document}

\title{Generating Binary Optimal Codes Using Heterogeneous Parallel Computing}
\author{\IEEEauthorblockN{Srajan Paliwal, Saurabh Tiwary, Bhaskar Chaudhury and Manish K. Gupta}
\IEEEauthorblockA{Group in Computational Science and High Performance Computing\\
DA-IICT, Gandhinagar, India\\
Email: bhaskar\_chaudhury@daiict.ac.in, mankg@computer.org}
}
\maketitle
\begin{abstract}
Generation of optimal codes is a well known problem in coding theory. Many computational approaches exist in the literature for finding record breaking codes. However generating codes with long lengths $n$ using serial algorithms is computationally very expensive, for example the worst case time complexity of a Greedy algorithm is $\mathcal{O}(n\; 4^n)$. In order to improve the efficiency of generating codes with long lengths, we propose and investigate some parallel algorithms using General Purpose Graphic Processing Units (GPGPU). This paper considers the implementation of parallel Greedy algorithm using GPGPU-CUDA (Computed Unified Device Architecture) framework and discusses various optimization techniques to accelerate the GPU code. The performance achieved for optimized parallel implementations is more than two to three orders of magnitude faster than that of serial implementation and shows a great potential of GPGPU in the field of coding theory applications.
\end{abstract}
\IEEEpeerreviewmaketitle
\section{Introduction}
\begin{figure}[h]
  \begin{center}
    \includegraphics[height=60mm, width=90mm]{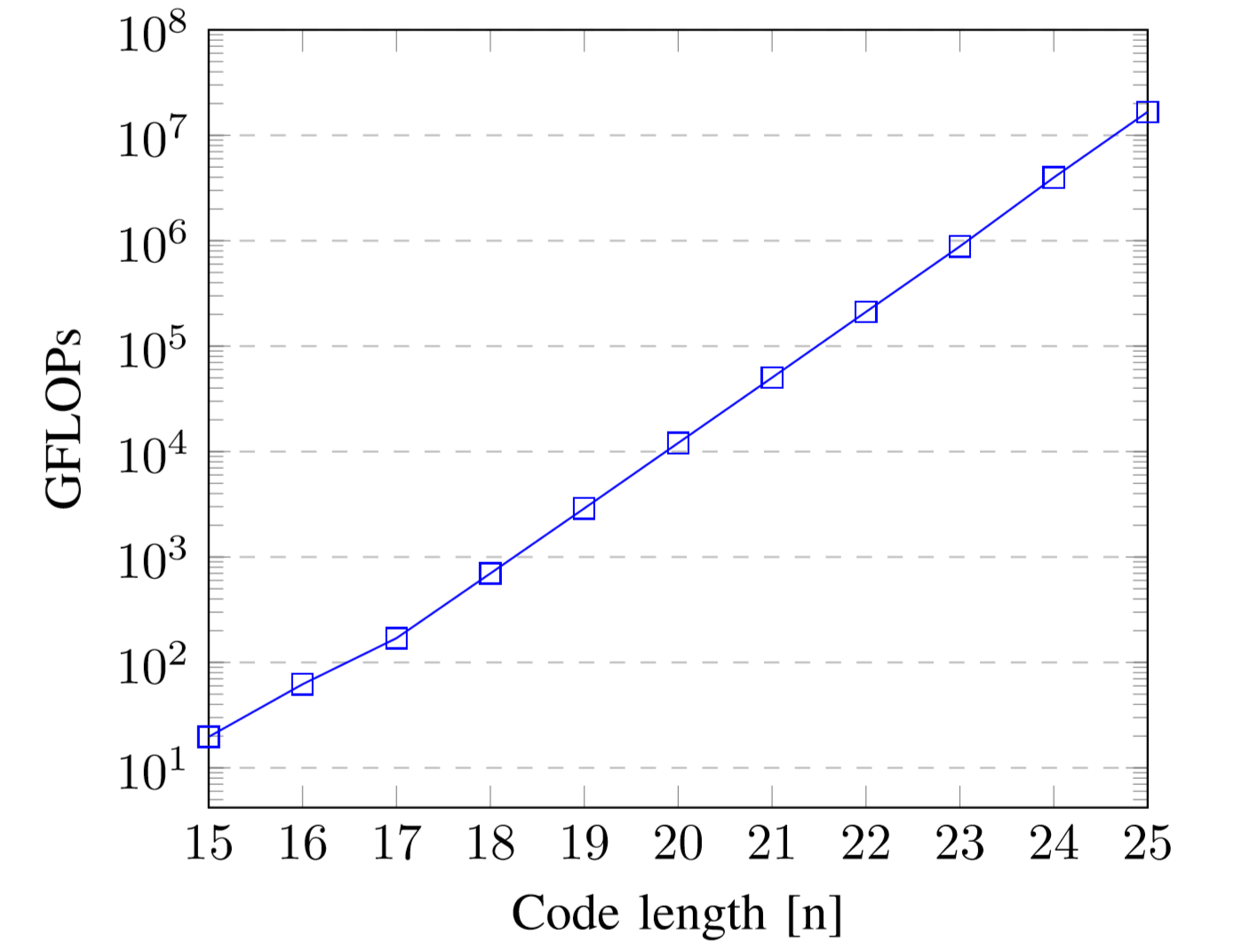}
    \caption{Computational demand (in GFLOPs) increases exponentially with increase in code length $\bf{n}$. This figure shows FLOPs required for various code lengths for minimum Hamming distance $d=3.$}
    \label{graph:GLOPSvscodelength}
  \end{center}
\end{figure}

Error correcting codes have applications in several fields such as communication, computer science, cryptography etc. Classification and generation of the optimal codes is a well known problem in coding theory (see for example \cite{Kaski:2005:CAC:1201556,Bouyukliev200151,Jaffe2000135}). 
In particular, computational approaches have resulted in many ground breaking codes \cite{Kaski:2005:CAC:1201556}. Since the 1990's, the beautiful greedy approach has been suggested and studied by many authors. This surprisingly produces many optimal codes by considering Hamming distance as a constraint \cite{journals/tit/ConwayS86a,Brualdi:1993:GC:173130.173134,FonDerFlaass1996156,Bonn:1996:FLG:235416.235419}. Notably, see \cite{conf/isit/GuendaGS12} for some recent work in this direction. It has been observed by many researchers that by changing ordering in the local choice, one can even produce optimal linear codes \cite{Bonn:1996:FLG:235416.235419}. In the review article \cite{acosta2001lexicographic}, four orderings have been discussed and many conjectures have been given for the construction of Greedy codes. The worst case time complexity of Greedy algorithm is $\mathcal{O}(n\; 4^n)$ with a serial computation, where $n$ is the length of the code. Hence for larger values of $n$ (See Figure \ref{graph:GLOPSvscodelength}) it is not feasible to get results quickly. When code length increases, required number of floating point operations (FLOPs) for code generation increases exponentially and is more than a few TeraFLOPS for "n" higher than 20 as shown in Figure~\ref{graph:GLOPSvscodelength}.

In the recent years, the emergence of GPGPU in scientific computing opens the way to low-cost massively parallel computations by using the large number of processors of a Graphics card to perform elementary calculations (such as calculation of Hamming distance) in parallel. A number of numerical tools for GPU computing have been developed in the last ten years. Furthermore, NVIDIA developed a programming environment called Compute Unified Device Architecture (CUDA) that allows creating efficient GPU codes and has become the foremost framework for GPGPU programming. CUDA utilizes the Single-instruction Multiple-threads (SIMT) architecture that enables a GPU to implement parallel data processing, and has become a parallel computing platform for different scientific applications.

The aim of this paper is to propose parallel algorithms for generating codes using GPGPUs, to measure the efficiency of these parallel algorithms compared to calculations on a single CPU, and to illustrate the methods with some examples. Our purpose is to give an overview of the possibilities and the constraints of programming a coding-theory algorithm on a GPU, and to provide an estimate of the gain in the computation time that can be obtained with respect to a standard CPU calculation. 
In this work, we have investigated the Greedy algorithm approach using GPGPU with different orderings like Lexicographic, Gray, Graded-lexicographic, Graded-reverse-Lexicographic as studied in \cite{acosta2001lexicographic}. In particular, we focus our attention for the construction of binary codes.

The paper is organized as follows. Section 2 gives a brief overview of GPU computing and CUDA environment. The basic background on coding theory and Greedy algorithms with different orderings are recalled in section 3. Parallel implementation of the Greedy algorithm is described in section 4. Several optimization techniques to improve the speedup are discussed in Section 5. Important results and some examples are presented in section 6 and 7 respectively. Finally we conclude the paper with general remarks.

\section{GPU Architecture and CUDA Environment}
The CUDA programming model is defined as a heterogeneous programming model which consists of C programming extensions with specially defined syntax. In this model, the serial part of the algorithm is executed on the host (CPU) side, and massively parallel data processing part is performed on the device (GPU) side. 

The number of technical books and papers devoted to GPU programming and CUDA environment is quickly increasing~\cite{pro_guid,best_pract_guid,kirk,lindholm}. We therefore limit ourselves to a brief discussion of parallel computing using GPUs in the CUDA environment. 
The GPU can be seen as a device composed of several thousands of cores (e.g., 2880 on the recent NVIDIA Tesla K40 GPU card used in this paper). 
The cores are grouped into multiprocessors (SMs), and each core can manage a set of concurrent threads. 
A thread is the basic programming unit of CUDA. A set of threads (Max. 1024 for Tesla K40 card) are organized to form a thread block to optimize the processor management, where they can synchronize their behavior and communicate. Threads are scheduled, managed, created and executed in a group of 32, called warp. All threads in a block reside in a multiprocessor and share its resources \cite{kirk}.

In the CUDA C environment, the programs that execute on the device side are defined as kernel functions which are executed N times by N different threads.  A kernel can be executed by more than several thousand threads.
The threads in one warp execute the same instruction simultaneously. Kernels can be launched from both CPUs or GPU. When a kernel is launched by a thead running on GPU, it is called Dynamic Parallelism \cite{pro_guid}.

An important aspect of developing efficient algorithms for GPUs involves good memory management. In a GPU, there are several types of memories: registers, shared memory, global memory, constant memory and texture memory. The variables invoked by kernel functions can be stored in either of the these five types of memory. Several different memory spaces can be accessed by threads on the device, but the access times of each memory type can differ by one or two orders of magnitudes. Typically, to access the global memory which is equivalent to the RAM on the CPU, a thread will take few hundred clock cycles, whereas it will take less than ten clock cycles to access the shared memory which is a 48-KB-size memory in the most recent GPU cards. Each block of threads has a shared memory, and all threads within the block can access this memory. The other two types of offchip memories, constant memory and texture memory, are cached and therefore the access speed of these two types of memory are much faster than that of global memory. Registers are allocated dynamically and privately to threads and provide most rapid access. The way the threads access the memory has a strong influence in kernel execution and latency~\cite{best_pract_guid,kirk}.
 Shared memory has low latency and limited capacity (presently few tens of KBs). 
The GPU uses a massive number of threads to hide the long latency of a global memory, instead of using a large cache as CPU. A CPU core is designed to minimize the latency for each thread at a time, whereas a GPU is designed to handle a large number of concurrent lightweight threads in order to maximize throughput.
 Unlike CUDA which can be implemented only on NVIDIA cards, Open Computing Language (OpenCL) is an open-standard environment that also allows GPU computing and can be run on any GPU hardware. However, CUDA is more mature, provides the best performance, and has more convenient high-level application programming interfaces. The implementation of OpenCL is close to that of CUDA C and the coding methods described in this paper can be easily adapted to OpenCL.
\section{Brief overview of the Greedy Algorithm and Codes}
A binary code $C$ of length $n$ is a subset of $\F^n_2$, where $\F_2$ is a binary field. If $C$ is also a subspace of dimension $k$, it is called a $[n,k]$ linear code. All the vectors in $C$ are known as codewords. One can define Hamming distance between any two code words as the number of places two codewords differ. Minimum Hamming distance among all codewords, denoted by $d$, is useful in the error detection and error correction when the code $C$ is used for some applications. If we define the weight of a codeword $c \in C$ as the number of non-zero places, then for a linear code $C$,
the minimum Hamming distance $d$ is minimum Hamming weight among all codewords of $C$. For a given $n$ and $d$, finding a code with the specific parameters is computationally  expensive problem and has been well studied by many researchers. If weights of each codeword in $C$ is constant, say, $w$ then such a code $C$ is called a code with constant weight $w$. In this paper, we focus our attention to the following problems. Firstly, given $n$, $d$ and $w$ construct the code $C$ using different orderings in greedy approach. Secondly, we focus our attention to the construction of Self orthogonal greedy codes. In particular we explore the search of class of extremely self dual codes $[24t, 12t,4t+ 4]$ which contains popular codes such as Golay code for $t=1$. 
\section{Greedy Algorithms and Different Orderings}
Given $n$, $d$ and $w$, we apply the Greedy algorithm to find a code $C$ with the given parameters. First step is ordering all the vectors of $\F^n_2$. Next, the vectors are appended to code $C$ from the set $\F^n_2$ one by one starting from zero vector. The vectors are checked to satisfy the constraint of minimum Hamming distance $d$ from all the previous choices in code $C$. The algorithm terminates when all the $2^n$ vectors are exhausted. In the case of self-orthogonal codes similar approach is taken.

\begin{figure}[h]
  \begin{center}
    \includegraphics[height=60mm, width=90mm]{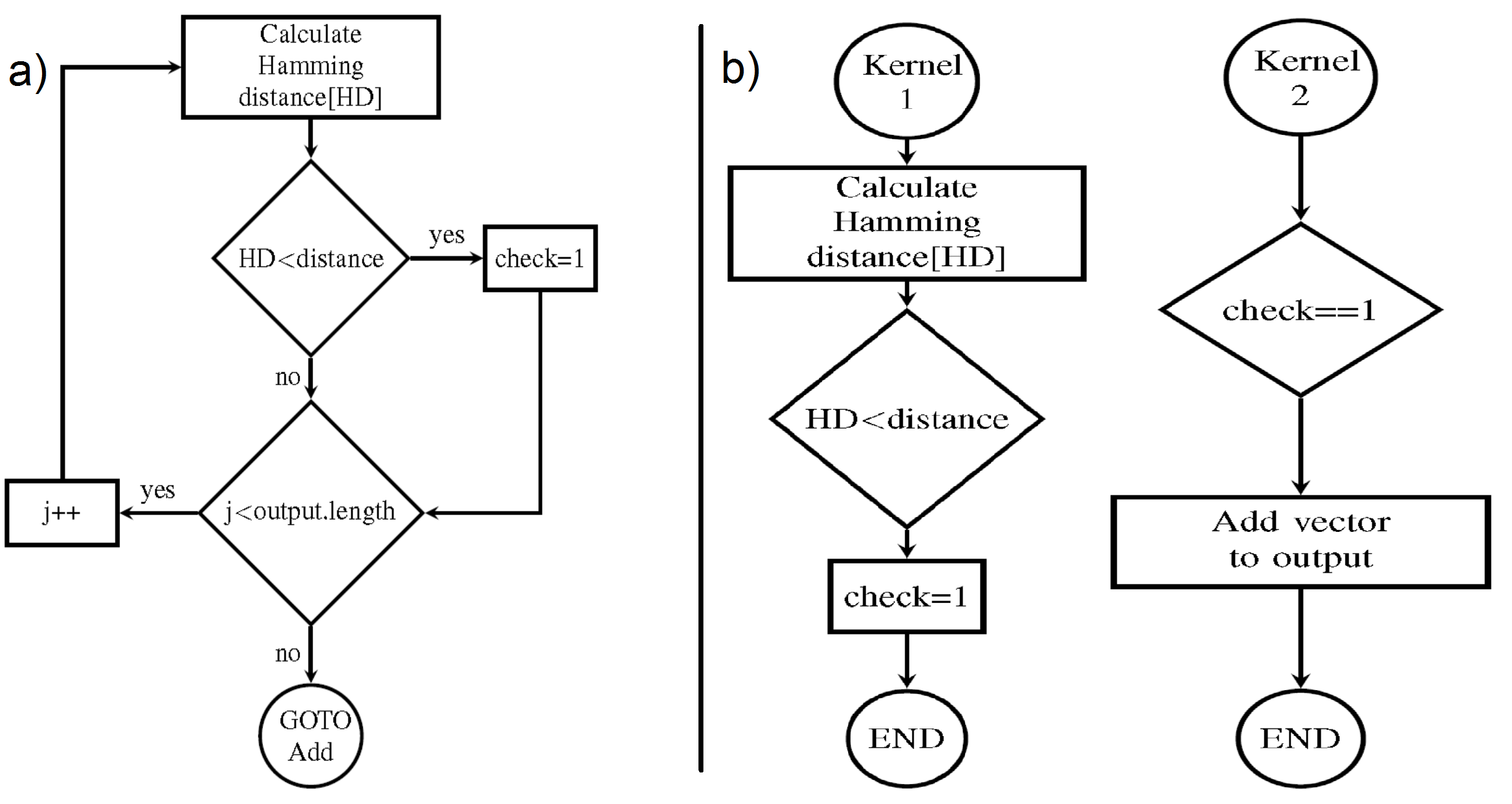}
\caption{(a) Serial implementation. (b) Parallel implementation}
\label{fig:SVP}
  \end{center}
\end{figure}

\subsection{Parallel Implementation} \label{Parallel Implementation}
For parallel implementation, the serial algorithm is modified into multiple independent sections that can run simultaneously on GPUs. As shown in Figure ~\ref{fig:SVP}, the inner loop in serial implementation calculates Hamming distances of current vector with all previously chosen code words. The algorithm performs this operation sequentially. However, the minimum Hamming distance condition check is independent for each code word. Thus, the minimum Hamming distance condition checks can be preformed concurrently. Figure~2(b) shows that the SIMT philosophy of GPU architecture has been used to divide the serial algorithm. Each section evaluates Hamming distance condition check, and sets the check flag based on the result.

\begin{figure}[h]
  \begin{center}
    \includegraphics[height=40mm, width=90mm]{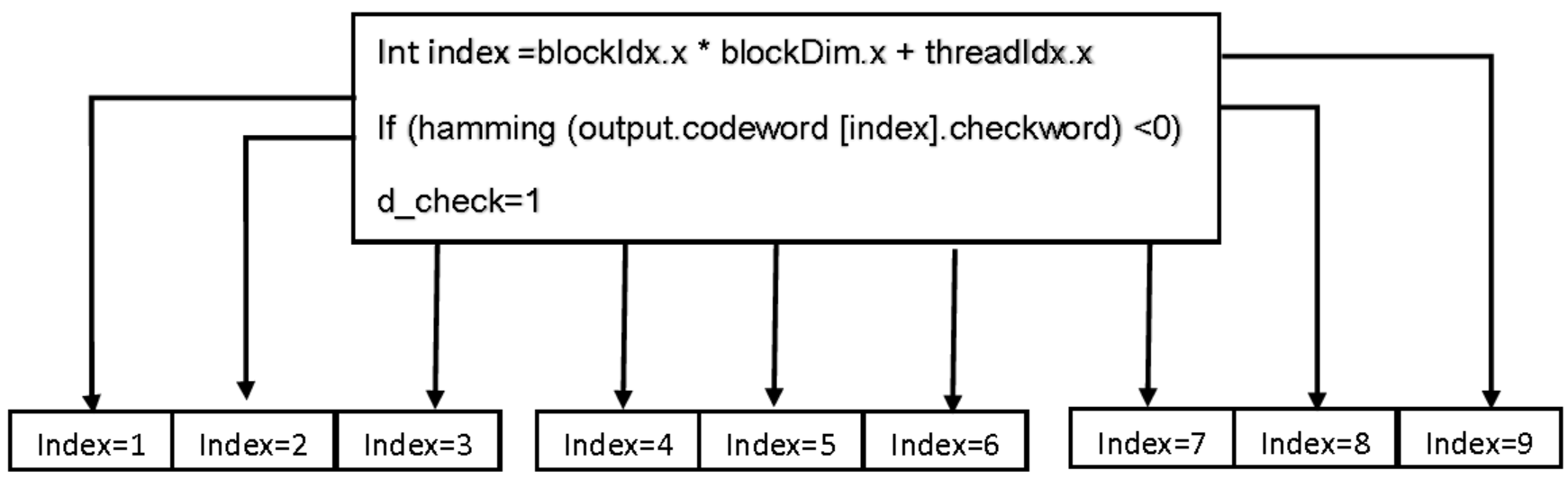}
\caption{Figure shows thread mapping on GPU multi-processors. The same kernel function is executed by each thread, but for different elements of output. The element with which Hamming distance is calculated is decided by the index. And, the index is given by blockIdx.x * blockDim.x + threadIdx.x. }
\label{fig:para}
  \end{center}
\end{figure}

The inner loop gets divided into sections equal to the length of current output. These sections are mapped to the various threads as shown in Figure~\ref{fig:para}. Each of these sections are executed by a different threads. The host launches two kernels. The first kernel calculates the Hamming distance and the second kernel adds that check word to the output code based on the check flag. Size of the output always gets copied from device to CPU memory before kernel launches for the next vector. Size of the output helps in deciding the grid and block dimensions for the next Hamming distance kernel.

First, the parallel algorithm allocates memory on the device corresponding to the size of output. Next, the  initialization kernel initializes the output with zero. Then, the host launches two kernel $2^n$ times for each vector in $\F^n_2$. Finally, the host copies the output code from device. Lot of overheads are attached with parallel computation like copying data and launching kernels. The parallel algorithm wastes time due to these overheads, which results in parallel algorithm taking more time than serial algorithm for smaller values of $n$.

In the following example~\ref{par}, we discuss in details the step by step implementation of a parallel algorithm. 

\begin{exmp}

Consider a code of length $n$=3 and distance $d \geq 2$. 
The input ordering of the eight vectors of the space is $\{000, 001, 010, 011, 100, 101, 110, 111\}$ which results in the output $\{000, 011, 101, 110\}$ using the following eight steps:
\\
Step 1 : Output is initialized with $000$.
\\
Step 2 : Output array size is checked and total number of threads are launched which equals the size of output array (size =1 till now). Vector $001$ is compared with $000$ and it does not satisfy minimum distance condition therefore it is not added to the final output. 
\\
Step 3 : Output array size is checked and total number of threads are launched which equals the size of the output array (size =1 till now). Vector 010 is compared with $000$ and it does not satisfy minimum distance condition therefore it is not added to the final output. 
\\ 
Step 4 : Output array size is checked and total numbers of threads are  launched which equals the size of the output array (size =1 till now). Vector $011$ is compared with $000$ and it satisfies minimum distance condition therefore it is added to the final output. 
\\
Step 5 : Output array size is checked and total number of threads are  launched which equals the size of the output array (size =2 now). Vector $100$ is compared with $000$ and $011$ and it does not satisfy minimum distance condition therefore it is not added to the final output. 
\\
Step 6: Output array size is checked and total number of threads are  launched which equals the size of the output array (size =2 till now). Vector $101$ is compared with $000$ and $011$ and it satisfies minimum distance condition therefore it is added to the final output.
\\
Step 7: Output array size is checked and total number of threads are  launched which equals the size of the output array (size =3 now). Vector $110$ is compared with $000$, $011$ and $101$ and it satisfies minimum distance condition, therefore it is added to the final output.
\\
Step 8: Output array size is checked and total number of threads are  launched which equals the size of the output array (size =4 now). Vector $111$ is compared with $000$, $011$, $101$ and $110$ and it does not satisfies minimum distance condition, therefore is not  added to the final output.
\\

\label{par}  
\end{exmp}

  \subsection{Ordering Implementation}
    The ordering of input vectors decides output codes. In this paper, four basic orderings are considered: Lexicographic order, Gray order, Graded Lexicographic order and Graded-Reverse Lexicographic order~\cite{acosta2001lexicographic}. Lexicographic order does not add any overhead to code calculation, as it is generated by iterating through $\F^n_2$. But for Graded Lexicographic order and Graded-Reverse Lexicographic order, the program selects all same weight vectors. These vectors are arranged based on the chosen graded ordering. The program executes the process of selecting same weight vectors by going through the vector space for every weight. Thus, the order generation is computationally expensive. B-ordering solves this problem of generating various orders.

\subsection{B-ordering Implementation}
A B-ordering is used for the construction of B-greedy codes~\cite{Brualdi:1993:GC:173130.173134}. Let $B=\{b_1,...,b_n\}$ be an ordered basis of $\F^n_2$. $B$ can induce an ordering of the vectors of $\F^n_2$ defined recursively as follows \cite{Brualdi:1993:GC:173130.173134}:
$\{0, b_1, b_2, b_2 + b_1, b_3, b_3 + b_1, b_3 + b_2, b_3 + b_2 + b_1, b_4, \ldots \}$. The recursive methodology looses its feasibility when used with parallel algorithm. Instead, a pair of nested loop replaces recursion, and produces vectors by adding basis elements corresponding to $1$'s place in lexicographic binary codes. Input basis decides the ordering of $2^n$ vectors. As described in the previous section, Graded Lexicographic order and Graded-Reverse Lexicographic order implementations generates codes by selecting all the weight and then arranging these vectors. Thus, the order generation take a larger portion of total computational time. B-ordering algorithm produce these orders by using their bases. B-ordering algorithm generates any orders in time which is equivalent to the time taken for generating a simple Lexicographic ordering. 

\subsection{Self Orthogonal Greedy Codes Implementation} 
A Self orthogonal greedy code~\cite{Monroe:1996:SGC:235416.235423} of length $n$ and distance $d$ is a Greedy code generated with the additional constraint that the vectors must be orthogonal to themselves and each other. The Hamming distance condition check kernel described in section \ref{Parallel Implementation} gets modified to include an additional check for orthogonality. First, the program AND the two vectors bit-wise. Then CUDA's built in function for population count counts the number of 1s. If the population count enumerates to a even number, then the vectors are orthogonal. The Hamming distance condition check kernel described in section \ref{Parallel Implementation} calculates Self orthogonal greedy codes.
\section{Optimization}
In the following sub-sections we briefly describe some of the optimization techniques which helps in achieving significant speedup over a naive parallel implementation.
\begin{figure}[h]
  \begin{center}
    \includegraphics[height=60mm, width=90mm]{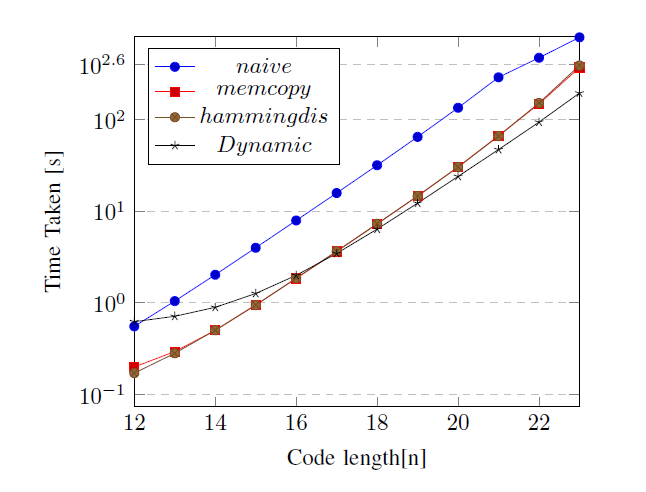}
    \caption{Time taken curves for each optimization step in
parallel implementation. Time taken curves are calculated for
minimum distance d=3 for different values of code length. In this figure - memcopy, hammingdis and dynamic represents the case $A$, $B$ and $D$ respectively and naive represents the case before removing wasted threads.}
    \label{optimisation:timings}
  \end{center}
\end{figure}

\subsection{Removing Wasted Threads} \label{Removing Wasted Threads}
Naive parallel algorithm launches two kernels. The first kernel calculates the Hamming distance, and the other kernel adds codewords to the final output. In the Hamming distance calculation kernel, the worst case scenario is taken into consideration. This results in wastage of large number of threads because of the initial assumption that all the codewords were added to the output.
   \begin{exmp}

For the same example as considered in previous section of code length 3 and distance at least 2 – 
\\
Step 1 : Considering vector $000$ is always added to the output.
\\
Step 2 : Considering vector $001$, one thread is launched.
\\
Step 3 : Considering vector $010$ and assuming the worst case scenario, two threads are launched assuming all the previous codewords were added to the output.
\\
Step 4 : Considering vector $110$. Three threads are launched, in this case assuming all other codewords were added. Two threads are getting wasted since only vector $000$ is there in the output.
\\
Similarly after 8 steps, seven threads will be launched, out of which 3 will be wasted because the output size is 4. For small length this is not a major problem, however as the code length increases, the number of threads wasted also increases, which consumes resources that could have been used for other useful computations.
\end{exmp}
To avoid this problem, every step is followed by copying of the output length  using “CUDAmemcopy” from device to host. According to this length, grid and block dimensions are calculated which in turn results in reduction of threads and thus prevents wastage of threads.
\subsection{Hamming Distance}
Hamming distance is equal to the number of places two code words differ. The naive way to implement this is to loop through the code length and compare each position. This method has a time complexity of $\mathcal{O}(n)$. But as we are working with binary codes, we can XOR two codewords that will give us the positions at which two codewords differ. Then, count the number of 1s. The result will give us the Hamming distance. CUDA has a fast population count function available and we used it to calculate the number of 1s. This method is not dependent on the length of the code. The time complexity is reduced from linear to constant time. In Figure ~\ref{optimisation:timings}, the graph of CUDAmemcopy optimization and constant time Hamming distance are overlapping because the minimum distance is small.
\subsection{Dynamic Parallelism}
Dynamic Parallelism makes it possible to launch kernels from threads running on the device. About 30\% to 35\% of the time in previous implementations was wasted in copying the current length of the output every time before launching a Hamming distance kernel. But with dynamic parallelism, the output size is available without CUDAmemcopy. The host launches a kernel with a single thread and then this thread  handles the rest of the kernel launches. Using dynamic kernel launch, one kernel can be launched inside another kernel. Thus, with the help off dynamic kernel launch, CUDAmemcopy part is removed. Figure~\ref{graph:speedupvscodelength} shows that it results in maximum increase in speedup.
\subsection{Selective Kernel Launch}
The program launches threads equal to the current size of the output as mentioned in section\ref{Removing Wasted Threads}. So, the increase in output size increase causes the number of threads to be launched to check each codeword to increase. A lot of computational power is wasted for every codeword that does not make it to output. So a program launching all the threads at once does not utilize the computational power of given GPGPU. Instead, this methodology evaluates restrains for initial 10 percent of the output. If the Hamming distance condition is satisfied, the kernel for rest of the threads is launched. First, the Hamming distance kernel with number of threads equivalent to the 10 percent of output length verifies the given restrains. If the check flag is not set after completion of this kernel, then the Hamming distance condition check kernel checks rest of the output. This reduces computational time a lot as total number of computations are reduced drastically, specially for large values of code length and distance.
\subsection{Recursion in Self orthogonal codes}
Greedy generation of self orthogonal codes, where $d$ is power of two, gives codes that can be utilized to generate codes with higher lengths~\cite{Monroe:1996:SGC:235416.235423}. The algorithm takes Cartesian product of two codes to produce higher length codes, where Cartesian product is the set of all ordered pairs such that codewords from second is appended to codewords of first set. So, the code$[n,d]$ can be produced using code $[n-k,d]$ and code$[k,d]$, where code$[k,d]$ is the repeating code for that value of $d$ or the length is a multiple of repeating code. By combining the previously described parallel algorithm with this method, codes of greater lengths can be constructed. This  algorithm achieves even higher speed-up. In Figure \ref{graph:selforthogonal}, we can see that it gives 3 order of speedups for higher values of n.
\begin{figure}[h]
  \begin{center}
    \includegraphics[height=60mm, width=90mm]{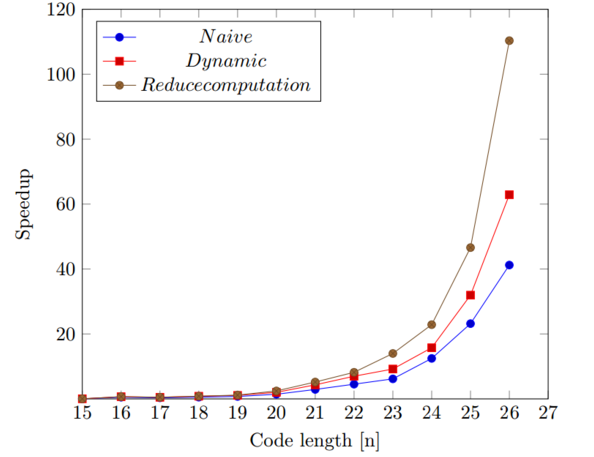}
    \caption{Speed-up vs code length for Hamming distance=$5$. In this figure, naive represents the case before removing wasted threads. Dynamic and reducecomputation represents the case $C$ and $D$ respectively.}
    \label{graph:speedupvscodelength}
  \end{center}
\end{figure}

\begin{figure}[h]
  \begin{center}
    \includegraphics[height=60mm, width=90mm]{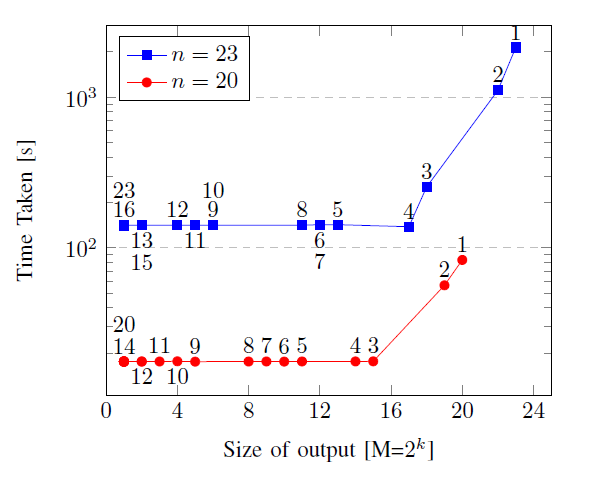}
    \caption{ Label at each coordinates show the given distance. The generated codes are linear as they are calculated using Lexicographic ordering. For linear code output size [M]=2k , where k is the dimension of this code. This graph show that, if size of output is not large enough , code is generated in constant time.
}
    \label{graph:distancevstimevscodelength}
  \end{center}
\end{figure}


\begin{table*}[t]
\def\arraystretch{1.5}
\begin{tabular}{|*{8}{p{1.8cm}|} }
\hline
Code Length & Serial Algorithm time (sec) & Parallel : Data Copy time (sec) & Percentage of total & Data Parallel Loop Time (sec) & Percentage of total time & Total Time  (sec) & Speedup \\ \hline
16          & 1.23        & 0.33                & 18.13               & 1.49                          & 81.87                    & 1.82              & 0.68    \\
18          & 5.18        & 0.33                & 5.23                & 5.98                          & 94.77                    & 6.31              & 0.82    \\
20          & 58.73       & 0.33                & 1.13                & 28.78                         & 98.87                    & 29.11             & 2.02    \\
21          & 248.15      & 0.33                & 0.58                & 57.05                         & 99.42                    & 57.38             & 4.32    \\
23          & 2093.42     & 0.33                & 0.15                & 227.24                        & 99.85                    & 227.57            & 9.20    \\
24          & 8711.30     & 0.33                & 0.06                & 552.73                        & 99.94                    & 553.06            & 15.75   \\
25          & 35609.67    & 0.33                & 0.04                & 763.87                        & 99.96                    & 764.20            & 46.60   \\
26          & 146221.06   & 0.33                & 0.02                & 1324.94                       & 99.98                    & 1325.27           & 110.33  \\ \hline
\end{tabular}
\label{table:cgm}
\caption{Speed-up for various values of n and Hamming distance= 5}
\label{table:cgma}
\end{table*}
\section{Results}

\begin{table}[h]
\def\arraystretch{1.5}
\resizebox{\columnwidth}{!}{%
\begin{tabular}{|l|l|l|l|l|}
\hline
\textbf{Card Name} & \textbf{Compute Capability} & \textbf{Cuda Core} & \textbf{Time Taken} \\ \hline
Tesla C2075 & 2 & 448 & 63.43 \\
Tesla K40c & 3.5  & 2880 & 98.05 \\
GTX 480 & 2 & 448 & 73.05 \\ \hline
\end{tabular}
}
\caption{Performance of various cards for $n=22$ and $d \geq 3$}
\label{table: deperform}
\end{table}

Figure~\ref{graph:speedupvscodelength} shows speedup for values of code length up to 26. For values up to 19, speed up is below 1. Codes of length up to 19 can be easily constructed on a CPU. If, we use a GPU to construct codes of smaller lengths (less than 20), we will be wasting available GPU resources. For even smaller values of n, GPU takes more time to generate codes. Major overheads associated with parallel computing are transferring data from host to device to work, allocating memory on the device or transferring processed data back to host. This leads to GPU taking more time than CPU to compute codes for smaller values of code length. This is clearly visible in Table~\ref{table:cgma}, where for smaller values on $n$, data copy time takes a major portion of total time. But as code length increases ($n$ greater than 18), the growth in speed up is exponential. Speedups only up to 26  were calculated because it is not feasible to run serial algorithm beyond that. But looking at the graph, we can predict that speedup will rise rapidly for higher value of code length. One important thing to notice in Table ~\ref{table:cgma} is percentage of data copy time in total time. For smaller values of n, it contributes almost half of the total time. But for a large value of n, its contribution is negligible and 99$\%$ of time is spent in code generation. This makes GPUs much more feasible to work on for this type of algorithms.  

We also calculated timings on different machines. Table ~\ref{table: deperform} has timing for generating code on length 22 and minimum distance 3 for different class of GPUs that were available to us. Tesla K40c took maximum time to compute codes due to lower memory bandwidth compared to the other two cards. Tesla K20 and GTX480 have very similar specifications, but increase in time from former to the latter is because of lower clock frequency for GTX480.

It is very important to understand when it is advisable to generate code using parallel computation and Figure ~\ref{graph:distancevstimevscodelength} clearly represents that. Time taken remains constant for large values of given minimum distance as well as the output size. If the given minimum distance is large compared to the code length, the number of comparisons required to calculate code decreases. Parallel algorithm does not scale well because of reduced number of comparisons.

\begin{figure}[h]
  \begin{center}
    \includegraphics[height=60mm, width=90mm]{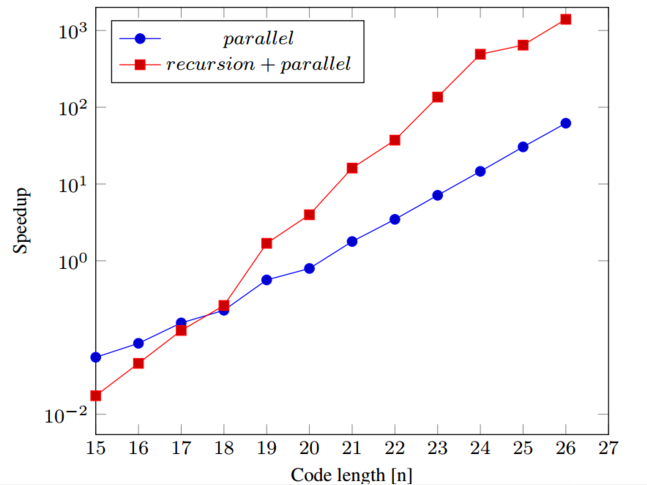}
    \caption{Comparing the speedup achieved with using parallel algorithm and recursion with parallel algorithm. This figure shows speedup calculated for d = 8 for n from 15 to 26.  }
    \label{graph:selforthogonal}
  \end{center}
\end{figure}

Self orthogonal greedy code ($n=8$,$d=4$) combined with Code$[4,4]$ produces Code$[12,4]$. All data obtained shows that cartesian product of Code(length-8,4) with Code(8,4) produces Codes of given length with distance equal to 4. The Figure \ref{graph:selforthogonal} compares speedup values of this method with only optimized parallel algorithm. For code length $n$ greater than 20, using this approach over only parallel is clearly much more viable. For n =26 , The algorithm achieved 3 order speedup which is 23 times more than optimized parallel algorithm. 
\section{Some Examples} \label{Some Examples}

We have proposed various parallel algorithm for generating codes using GPGPUs and different optimization techniques associated with each algorithm. The optimised parallel algorithm  gives two order speedup compared to the serial algorithm. Most of the optimisation techniques discussed here emphasized on reducing computational time of Greedy algorithm itself. This enables easily integrating other modules into the core Greedy algorithm, one such example investigated here was self orthogonal codes. Now, we discuss few other examples that demonstrate efficiency of optimized parallel algorithm over serial algorithm.

One of the cases observed in the Lexicographic and Non-Lexicographic greedy codes~\cite{acosta2001lexicographic} was that the codes generated for n = 7 and d = 3, also for n = 15 and d = 3 were perfect codes. But, they were not able to confirm this for n = 23 and d = 7 due to computational limitations, however other researchers have shown that Golay code is a lexicode~\cite{Brualdi:1993:GC:173130.173134}. However with our parallel implementation, we can easily generate code for n=23 and d=7, and reconfirmed that generated code is Golay code with n = 23, k= 12 and d = 7.

In~\cite{acosta2001lexicographic}, a conjecture regarding the Lexicographic and Non-Lexicographic greedy codes was proposed, but due to computational limitations authors were not able to confirm them for higher values of n. If the Hamming distance d is a power of 2, the Greedy codes generated with the lexicographic order and the graded lexicographic order will contain exactly the same words. They were able to observe this only till n = 15 and d = 8. We were able to generate codes up to n = 30 for d = 16 and observed the same trend.

In~\cite{Monroe:1996:SGC:235416.235423} Laura Monroe proposed a conjecture that with Lexicographic ordering for any distance, greedy generation of self orthogonal codes will produce self dual code. Using CRAY C-90, they were able to produce codes till baby Golay code $[22,11,6]$ and extended Golay code $[24,12,8]$. We were able to verify this conjecture for code length $n=30$ and distance $d=10,15,20$ very easily.

The existence of a $[24k, 12k, 4k + 4]$ Type II self-dual code is one of the most famous open problems when k = 3 (see~\cite{dougherty2011search} for details on this problem). The greedy generation of self orthogonal code produces extended Golay code $[24,12,8]$. This motivates us to look at the next interesting cases of producing Type II self-dual codes using parallel algorithms for code length 48 and 72. Various possibilities and optimisation techniques presented in this paper combined with other algorithms for code generation may result in code with code length as high as 72. 

We have also computed binary constant weight codes using parallel approaches described in the paper and verified the published results on constant weight codes of length up-to 35. 


\section{Conclusions}
The paper proposes several parallel algorithms for generating codes using GPGPUs and presents a measure of the efficiency of these parallel algorithms compared to calculations on a single CPU. In this work, we have investigated the Greedy algorithm approach using GPGPUs with different orderings like Lexicographic, Gray, Graded-lexicographic, Graded-reverse-Lexicographic as studied in \cite{acosta2001lexicographic}. The optimized parallel algorithms for generating codes on GPGPUs described in this paper results in more than 2 orders of magnitude speedup compared to a serial CPU code. This makes it feasible to generate codes of length greater than 23 in a computationally much less expensive manner. The self orthogonal greedy code construction using parallel algorithm performed much better due to recursive nature and 3 orders of speedup was achieved.  Two conjectures, for self orthogonal greedy code and Lexicographic greedy codes, were verified for higher values of code length as described in literature. Finally, we have illustrated the methods with some useful examples. The work gives an overview of the possibilities and the constraints of programming a coding-theory problem on a GPU, and provide a quantitative  estimate of the gain in the computation time that can be obtained with respect to a standard CPU calculation. To the best of our knowledge, the techniques described in this work will be a first step towards using GPGPU approaches for solving other popular problems in coding theory  such as computing the covering radius of a code, finding the generalized Hamming weight or even decoding a code. 

\section*{Acknowledgments}
The authors would like to thank the NSF-supported Center for Parallel and Distributed Computing Curriculum Development and Educational Resources (CDER) center at Georgia State University for providing the parallel computation facilities.
\bibliography{HPCcodes}
\bibliographystyle{plain}
\end{document}